\documentclass[prd,twocolumn,floatfix,preprintnumbers,letterpaper]{revtex4}
\usepackage{graphicx}
\usepackage{amsmath,amssymb}
\input{epsf}
\usepackage{epsf}
\newcommand {\ga} {\ {\raise-.5ex\hbox{$\buildrel>\over\sim$}}\ }
\newcommand {\la} {\ {\raise-.5ex\hbox{$\buildrel<\over\sim$}}\ }

\begin{document}

\def\be{\begin{equation}}
\def\ee{\end{equation}}

\title{Radiation can never again dominate Matter in a Vacuum Dominated Universe }
\author{Lawrence M. Krauss$^{1,2}$ and Robert J. Scherrer$^{2}$}
\affiliation{$^1$CERCA, Department of Physics, Case Western Reserve University,
Cleveland, OH~~44106}
\affiliation{$^2$Department of Physics \& Astronomy, Vanderbilt University,
Nashville, TN~~37235}
\date{\today}

\begin{abstract}
We demonstrate that in a vacuum-energy-dominated expansion
phase, surprisingly neither the decay of matter nor matter-antimatter annihilation into relativistic
particles can ever cause radiation to once again dominate over matter in the future history
of the universe.
\end{abstract}

\maketitle

The study of decaying particles in a cosmological context has a long
history \cite{Lindley,Wein,TSK,ST1,ST2,ST3,LMK1,LMK2,LMK3,Protondecay}. 
Although there are many variations
on this theme, there is one constant result:  if the universe is initially
dominated by nonrelativistic particles, and this matter undergoes a standard
exponential decay into relativistic particles, then the universe
rapidly transitions from a matter-dominated phase into a radiation-dominated
phase when the age of the universe is roughly equal to the particle
lifetime.

In this paper we show, rather suprisingly, that this conclusion is no
longer valid at the present time, if the dark energy reflects a constant vacuum energy density in the universe.  Once the universe begins to enter a phase of exponential
expansion, no process leading to the disappearance of matter can ever cause radiation
to once again dominate over the remnant matter density, no matter how small.

Consider a nonrelativistic component with density $\rho_M$, decaying
into a relativistic component $\rho_R$, and take the lifetime
of this decay to be $\tau$, with a standard
exponential decay law.  (For simplicity, we will assume
that the decay products are ``sterile", i.e., that they do not
interact significantly with anything else, but our conclusions do not
depend on this assumption).  Then the equations governing the matter
and radiation evolution are
\begin{eqnarray}
\label{decay1}
\frac{d\rho_M}{dt} &=& -3H \rho_M - \rho_M/\tau,\\
\label{decay2}
\frac{d\rho_R}{dt} &=& -4H \rho_R + \rho_M/\tau,
\end{eqnarray}
where $H$ is the time-dependent Hubble parameter:
\begin{equation}
H = \left(\frac{8 \pi G \rho}{3}\right)^{1/2},
\end{equation}
with $\rho$ being the total energy density.
(We assume a flat universe throughout).
If the decaying nonrelativistic component dominates the expansion,
it is easy to show that the
energy density in the nonrelativistic component is rapidly
converted into relativistic energy density when $t \approx \tau$
(see, e.g., Ref. \cite{ST1}); indeed, this assumption has become
part of the standard cosmological lore.

But now consider what happens for a vacuum-dominated expansion,
such as the universe is experiencing at present.  We
take the ratio of radiation density to matter density to be
given by $r$:
\begin{equation}
r \equiv \rho_R/\rho_M.
\end{equation}
Then equations (\ref{decay1})$-$(\ref{decay2}) can be combined
to yield an equation for $r$: 
\begin{equation}
\label{r}
\frac{dr}{dt} = \frac{1}{\tau} + \left(\frac{1}{\tau} - H\right) r.
\end{equation}
In the matter-dominated era, $H$ decreases with time, and as
long as $H < 1/\tau$, we see that $r \rightarrow \infty$, as expected.
However, as the universe evolves from a matter-dominated state to a vacuum-energy
dominated state, the value of $H$ asymptotically approaches a constant
value, $H_{\Lambda}$, given by
\begin{equation}
H_{\Lambda} = \left(\frac{8 \pi G \rho_{\Lambda}}{3}\right)^{1/2},
\end{equation}
where $\rho_\Lambda$ is the (constant) vacuum energy density.
Define a `time of no return',  $t_{\Lambda}  \equiv 1/H_\Lambda$.  
Substituting $H = H_{\Lambda} \equiv 1/t_\Lambda$ into equation (\ref{r}), we can solve
this equation analytically to yield
\begin{equation}
\label{solution}
 \left(1 - \frac{\tau}{t_\Lambda}\right) r = \exp\left(\left[\frac{1}{\tau}
- \frac{1}{t_\Lambda}\right]t\right) - 1,
\end{equation}
where $r$ is normalized to be zero at $t=0$, i.e. the initial radiation density is negligible in the  matter-dominated era.

In terms
of the present-day Hubble parameter $H_0$, and the fraction of 
the critical density in vacuum energy, $\Omega_\Lambda$, we have
simply
\begin{equation}
\label{tL}
t_{\Lambda} = H_0^{-1} \Omega_{\Lambda}^{-1/2} = 9.8 \times 10^9 ~{\rm yr}
~ h^{-1} \Omega_\Lambda^{-1/2},
\end{equation}
where $h$ is the value of $H_0$ in units of 100 km sec$^{-1}$ Mpc$^{-1}$.

\begin{figure}[t]
\centerline{\epsfxsize=3.7truein\epsfbox{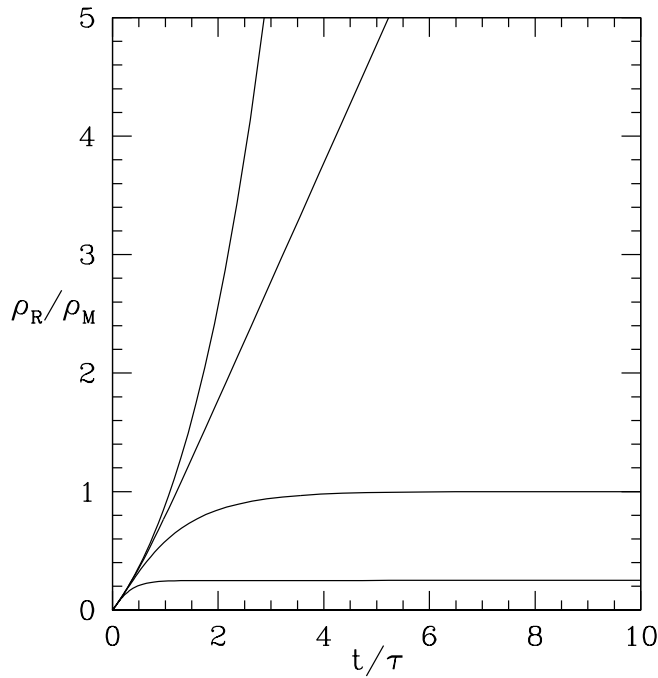}}
\caption{The ratio of decay-produced relativistic energy density,
$\rho_R$, to decaying nonrelativistic dark matter, $\rho_M$,
as a function of the time measured in units of the decaying
particle lifetime $\tau$.  From top to bottom, the curves
correspond to $\tau/t_\Lambda = 0.5, 1, 2, 5$.}
\end{figure}

The asymptotic value of $r$ is sensitive to the relative values of
$\tau$ and $t_\Lambda$.  If $\tau < t_\Lambda$, then as $t \rightarrow
\infty$, equation (\ref{solution}) gives $r \rightarrow \infty$,
which is the same as the conventional result for decays during the
matter-dominated phase, since at such times the vacuum energy will indeed
be cosmologically irrelevant.  But if $\tau > t_\Lambda$, the right-hand side
of equation (\ref{solution}) goes to $-1$ as $t \rightarrow \infty$,
and $r$ asymptotically approaches a constant, given by
\begin{equation}
r(t\rightarrow \infty) = \frac{t_\Lambda}{\tau - t_\Lambda}.
\end{equation}
Thus, the ratio of radiation to matter approaches a constant; the matter
never disappears relative to the radiation (although of course
both densities go to zero as $t \rightarrow \infty$).  Further,
as long as $\tau > 2 t_\Lambda$, we have $r < 1$ asymptotically;
in this case the decay-produced radiation never dominates the decaying
matter!  Finally, there is an intermediate case, $\tau = t_\Lambda$,
for which the solution given in equation (\ref{solution}) does not
apply.  In the case, the solution of the evolution equation is simply
$r = t/\tau$.  In this case,
$r$ becomes arbitrarily large as $t \rightarrow \infty$, but the
increase is linear in time, rather than exponential as is the case
for shorter particle lifetimes. (See also \cite{barrow} which examines a different model for energy exchange).

While this result is perhaps non-intuitive, there is a straightforward physical explanation.
Radiation redshifts as one extra power of the scale factor
relative to matter.  For a vacuum energy-dominated universe the
scale factor is itself an exponential of the time. Thus, in the absence of decay the energy density
of matter in the universe will increase relative to radiation by an exponential function of time.  If the matter
density itself decreases exponentially, the two factors cancel, leading to a constant final
ratio of matter density to radiation density.

Our analytic solutions give the correct asymptotic behavior,
but equation (\ref{r}) must be integrated numerically to study the
physically realistic case of a universe containing both decaying dark matter
and vacuum energy.  We have performed this numerical
integration for several illustrative values of $\tau/t_\Lambda$, (note that
$\rho_M/ \rho_{\Lambda}$ at any time is a fixed function of $t/t_{\Lambda}$); the
results are displayed in Fig. 1.
This figure clearly shows the asymptotic behavior discussed above.

Now consider our present-day universe.  Taking reasonable values
of $h = 0.7$ and $\Omega_\Lambda = 0.7$ in equation (\ref{tL}), we
obtain a value of $t_\Lambda = 1.7 \times 10^{10}$ yr.  Current constraints
on the decay of the dark matter yield $\tau > t_\Lambda$.  For
example, Ichiki et al. \cite{Ichiki}, using the WMAP data, derive
a 95\% confidence limit of $\tau > 5.2 \times 10^{10}$ yr,
which corresponds to $\tau/t_\Lambda > 3$.  Thus, we conclude
that we have already entered the era at which it is no longer possible
for relativistic energy density from decaying dark matter to ever
dominate the dark matter density itself.

There are other mechanisms to convert the dark matter
into relativistic energy density.  For example 
dark matter within the dense cores of halos can annihilate into relativistic particles (see, e.g.,
Ref. \cite{Kaplinghat,Zentner}).  As Krauss and Starkman
have recently demonstrated \cite{ann}, such processes will have dramatic consequences for the future of large scale structure.

In general, as described in \cite{ann}, at late times the annihilation rate for dark matter in halos with
density $n$, given by $\Gamma_{ANN} = n \langle \sigma v \rangle$, will fall as $t^{-1}$.    Nevertheless, even if it were to remain constant over time, this annihilation rate would still
not be fast enough produce a radiation density that would overwhelm that in matter.   

In the case of falling annihilation rates one can in fact produce a stronger bound.   If $\dot{M} =M/t$, which will be the case for bound systems whose density changes only due to annihilations as the universe ages  \cite{chibaadams}, then $M(t) =M_0/t$.  In this case, as long as the expansion of the universe is such that $R \approx t^{\alpha}$ where $\alpha \ge 1$ then radiation will redshift faster than the mass decreases, and the radiation density will never overwhelm the matter density.  This will occur if the equation of state parameter $w =p/\rho < -1/3$ for the dominant energy in the universe.    However, for systems bound by gravity, where adiabatic expansion will accompany annihilations, supplanting annihilation in reducing core densities, then $\dot{M} =M/4t$ \cite{ann}.  In this case, $M(t) =M_0/t^{1/4}$.   For this case, no expansion involving radiation or matter is sufficiently slow to allow the radiation produced by annihilations to overwhelm the matter density.
  
In the unphysical but conservative case where annihilation occurs at a constant rate,  
dark matter annihilation would then map directly
onto the decaying particle problem we have already solved.
With a constant annihilation rate, each galaxy halo would act as a
very massive decaying ``particle",
with lifetime $\tau = 1/\Gamma_{ANN}$, and our previous analysis
would then hold.  
Studies of galaxy halo profiles today (i.e.  \cite{Kaplinghat,Beacom}) 
imply that the characteristic annihilation time, $\tau$, in galaxies must greatly
exceed the present age of the universe, and hence also greatly
exceed $t_{\Lambda}$ today.  Moreover for annihilation into standard particles
and for dark matter whose remnant abundance today was determined by
freezeout at early times, the inverse annihilation rate will in fact be many orders
of magnitude larger than the current age today.  In either case,
our previous argument holds:  the ratio of $\rho_R$ to $\rho_M$ will
remain small regardless of the particle annihilations.

Our consideration of dark matter annihilation is particularly relevant if
ordinary matter is in fact unstable, due to proton decay.  In this case, our arguments
tell us that even if dark matter annihilates away, remnant baryons will still dominate compared to radiation.  Moreover, even once baryons start to decay significantly in the far, far future \cite{dic} they will still dominate over the energy density of their relativistic decay products until these decay products themselves become non-relativistic, assuming they are massive.   Thus, matter, even ordinary matter, will always continue to dominate over radiation for all times.

We further point out in this regard that  our analysis clearly applies in the case of two decaying matter species with different timescales, as for example might occur for decaying dark matter and decaying baryons.  Since the latter are likely to have a far longer lifetime, our analysis is trivially extended by considering two species with different exponential timescales for decay.  Consider two such
species with lifetimes $\tau_1$ and $\tau_2$, with
$\tau_1 << \tau_2$.  Once $t > \tau_1$, while the ratio of energy densities
in the first decaying particle and the radiation produced by it will become constant, nevertheless the total energy
density in both will quickly become insignificant relative to the energy
density in the second, more stable, nonrelativistic species.  Then when the second
nonrelativistic species decays, our original analysis applies. 

In the extremely far future, individual galaxies, or clusters of galaxies will become essentially isolated `island universes' \cite{island}.  Nevertheless, on sufficiently large (perhaps even super-horizon) scales these will behave like a homogenous gas of particles amidst a background vacuum energy sea from the perspective of this analysis, and if the matter within them decays or annihilates, as discussed above, they can be treated as decaying particles and our analysis continues to hold.

To be sure, it is true that in the far future of a vacuum-dominated universe, neither matter nor radiation will significantly affect the dynamics of the universe, and therefore which dominates over  the other is, from a practical perspective, not dynamically significant.  Nevertheless, many scenarios in early universe cosmology are based on decaying systems which cause the universe to shift from matter to radiation domination.  One's intuition suggests this will always be the case, and thus this result is interesting because it demonstrates that conventional wisdom about the past universe cannot be applied to the future universe, if it remains vacuum or dark energy dominated. In this regard, our result may also be relevant for determining the initial conditions of a post-vacuum dominated universe, if the dark energy is not vacuum energy, but instead evolves. 

This latter comment, of course, reinforces the fact that our results depend crucially on the assumption that the
dark energy is, indeed, a vacuum energy (cosmological constant)
with $\rho_\Lambda = constant$, and that it, itself, does not decrease with time, in which
case the Hubble parameter will once again decrease with time.   Thus, in spite of
the many negative facets of a future with a cosmological constant \cite{fut}, only persistent vacuum energy
remarkably preserves this surprising eternal dominance of matter over radiation.

\acknowledgements 
L.M.K. was supported in part by a Department of Energy grant and by an NASA  Astrophysics Theory grant, and thanks Vanderbilt University for hospitality while this work was carried out.
R.J.S. was supported in part by the Department of Energy (DE-FG05-85ER40226).

\end{document}